\documentclass[12pt]{article}

\usepackage{newtxtext,newtxmath}

\usepackage{graphicx}

\usepackage[letterpaper,margin=1in]{geometry}

\renewenvironment{abstract}
	{\quotation}
	{\endquotation}

\date{}

\makeatletter
\renewcommand{\fnum@figure}{\textbf{Figure \thefigure}}
\renewcommand{\fnum@table}{\textbf{Table \thetable}}
\makeatother

\usepackage{amsmath}
\usepackage{url}

\usepackage{scicite}

\usepackage[table]{xcolor}
\definecolor{lightgray}{rgb}{0.9,0.9,0.9}
\usepackage{array}
\usepackage{setspace}
\newcolumntype{C}[1]{>{\centering\arraybackslash}m{#1}}

\makeatletter
\let\old@cite\cite
\renewcommand\cite{\protect\old@cite}
\makeatother

\def\shorttitle{Survivable Niches for Microbial Life on the Lunar South Pole}

\def\teaser{Multiple microbial species might be able to survive in portions of lunar polar regions relevant to exploration and habitation}

\def\scititle{
	Potential survivable niches for microbial life on the lunar south pole
}
\title{\bfseries \boldmath \scititle}

\author{
	Prabal~Saxena$^{1\ast}$,
	Stefano~Bertone$^{1,2,5}$,
	Heather~V.~Graham$^{1}$,
	Natalie~M.~Curran$^{1,3}$\and
	Aaron~B.~Regberg$^{4}$,
	Andrew~Needham$^{1}$,
	D.E.~(Betsy)~Pugel$^{1}$,
	Noah~E.~Petro$^{1}$\and
	\small$^{1}$NASA Goddard Space Flight Center, Greenbelt, MD 20771, USA.\and
	\small$^{2}$University of Maryland, CRESST II, College Park, MD 20742, USA.\and
	\small$^{3}$Catholic University of America, CRESST II, Washington, DC 20064, USA.\and
	\small$^{4}$NASA Johnson Space Center, Houston, TX 77058, USA.\and
	\small$^{5}$National Institute for Astrophysics (INAF), Astrophysical Observatory of Turin, 10025 Pino Torinese, Italy.\and
	\small$^\ast$Corresponding author. Email: prabal.saxena@nasa.gov
}

\begin{document} 

\maketitle

\noindent\textbf{Short title :} \shorttitle\\
\textbf{Teaser sentence:} \teaser
\begin{abstract} \bfseries \boldmath
Most lunar surface conditions are incredibly harsh for microbial survival. High ultraviolet radiation, temperatures, and energetic particle radiation limit survival over most unprotected lunar surfaces, particularly in equatorial regions where all previous crewed exploration occurred. However, whether these harsh conditions are widespread at lunar poles has not been examined considering topographical effects. Here, we show that recent microorganism survivability data and lunar surface remote sensing reveal likely survivable niches in lunar polar regions. Analysis of topography and latitude-driven surface conditions using remote sensing data and high-resolution illumination models indicates the lunar south pole possesses significant regions with persistent low temperatures and ultraviolet flux. Comparing these conditions to survivability data of specific microorganisms, we find significant lunar polar areas likely possess surface conditions amenable to microbial survival. Our findings suggest lunar polar regions may be less hostile to microbial survival than previously assumed. This does not encompass growth likelihood, but survival in a cryptobiotic state where growth would be possible if habitable conditions were present. Potential microbial survivability at lunar poles is particularly significant given many examined microbes will likely be transported to the Moon during crewed lunar south pole exploration planned in numerous near-term missions. Thoughtfully planning exploration and tracking its impact is key to limiting and understanding potential unintended life transfer to the Moon.
\end{abstract}


\section*{Introduction}

The near term plans for a sustained surface presence at the lunar south pole represent a significant shift in human space exploration. This unprecedented plan for habitation at a location outside low earth orbit offers unique opportunities for science, exploration, resource identification and extraction, and engagement of the general public with space. However, such plans would also inevitably lead to an anthropogenic influence on the environment surrounding the habitat and exploration zones, thus affecting scientific questions including those related to prebiotic molecules and other volatile organic compounds \cite{2024BAAS...56a.009K, 2022SpReT.215...45C} that can be posed at these locations.

The lunar south pole possesses unique properties relative to the rest of the Moon, some of which enable the presence of surface/near-surface water ice, which is part of the motivation for the heightened interest in the region. Low Sun elevation angles at the lunar south pole due to the Moon's orbit and obliquity and to regional topography, result in portions of the surface that receive significantly attenuated amounts of solar radiation. These factors are what enables the presence and stability of water ice\cite{watson1961possible, arnold1979ice, 2020PSJ.....1...54S} and generally lead to low amounts of Ultra-Violet (UV) light, low temperatures, and low energetic particle flux in regions at the south pole relative to the rest of the Moon.

These unique properties present an interesting test case with respect to the survival of life - the present day Moon and exposed space have generally been regarded as extremely harsh environments for microbial life \cite{2000ASPC..213..593H, 2008AcAau..63.1006K, doi:10.1089/ast.2018.1952}. A previous study\cite{doi:10.1089/ast.2018.1952} found that UV light exposure and high temperatures are the key factors in making space environments the most hostile for microbial life, and that the likelihood of survival of microbial life left on the lunar surface by previous exploration in equatorial regions of the Moon was deemed to be low. However, the model did not include the effects of topography; while these effects may be muted at regions with potentially high Sun elevation angles (such as equatorial regions), in polar regions topography plays a significant role in local conditions relevant to microbial survival.

This is particularly important given the context of the evolving understanding of the survival of microbes exposed to space environments. Microbial survivability in space environments is not merely conditional on the external stresses it is exposed to, but also to the microorganism's response mechanisms including structural, metabolic, genetic and even ecological features\cite{doi:10.1128/MMBR.00016-09, 10.3389/fmicb.2019.00780, 10.3389/fmicb.2020.00923}. Intentional and serendipitous space exposure\cite{HORNECK199441, Ott2020, https://doi.org/10.1111/1462-2920.15995, Deshevaya2020, Deshevaya2022} and laboratory experiments\cite{doi:10.1128/AEM.03007-06, 10.3389/fmicb.2020.00560, doi:10.1089/ast.2019.2214} suggest an evolving understanding of microbial survivability in space. The potential that certain microbes may be capable of surviving space conditions that may be relevant to the lunar south pole motivates a joint analysis by our team of observed/inferred surface conditions at the lunar south pole versus survival bounds of relevant microbes.

We focus on three bacterial genera and two fungal genera; four that are representative of the ubiquitous of opportunistic microbes likely to be transported to the lunar surface as part of human exploration, as well as an extremophilic bacteria that has been shown to be particularly well adapted to the harsh conditions of space environments and has been cultured in NASA clean rooms. We consider survival bounds of species within these genera to determine a plausible phase space for the two most critical biocidal factors on the lunar surface - UV radiation and temperature. These genera are resistant to vacuum conditions\cite{doi:10.1128/MMBR.00016-09, 10.3389/fmicb.2019.00780,HORNECK199441,doi:10.1089/ast.2018.1952,doi:10.1089/ast.2018.1913,DOSE1995119, SARANTOPOULOU201163}, temperature extremes, and energetic particle radiation, with specific discussion of the examined microbes' survivability under these conditions given in supplementary materials. Importantly, these are all microbes that are commonly found in crewed space habitats\cite{10.3389/fmicb.2021.608478, doi:10.1089/ast.2017.1751, doi:10.1089/ast.2018.1913, OLSSONFRANCIS20101, VESPER2008432, doi:10.1089/ast.2020.2399, doi:10.1089/ast.2021.0049, https://doi.org/10.1111/j.1365-2672.1996.tb03541.x, doi:10.1089/ast.2018.1854, onyango2012effect}, which makes them more likely to be transferred to the lunar surface given that such a habitat would accompany human exploration of the surface. Since there is a well established correlation between solar sourced particle radiation and UV radiation\cite{doi:10.1089/ast.2018.1952}, we use UV exposure and temperature as proxies to map local conditions at the lunar poles and identify specific locations where microorganisms could survive for at least one Earth day. This timescale is chosen because one Earth day is greater than the longest time between consecutive extravehicular activities (EVA's) previously conducted on the moon. Microbial survivability between consecutive EVAs has the potential to impact scientific operations by increasing the organic contamination baseline in Permanently Shadowed Regions (PSRs)\cite{witze2021will} and confounding our search for prebiotic molecules and other volatile organic compounds that may be concentrated in these locations\cite{LawrencePolarVolatiles, LUCEY2022125858}.

This survival estimate leverages recent remote sensing products, which allowed us to characterize surface conditions at highly resolved spatial scales, allowing us to consider the shadowing effects of local topography. Temperature data is based on measurements by Lunar Reconaissance Orbiter's (LRO) Diviner\cite{https://doi.org/10.1029/2019JE006028} that yields seasonal temperatures at the scale of 240 meters, while regional UV ($\leq$ 320nm) fluxes can be estimated based on average illumination maps with a pixel scale of 60 meters based on topography derived from LRO Lunar Orbiter Laser Altimeter (LOLA) measurements\cite{MAZARICO20111066}. We then analyzed UV fluxes at finer spatial scales for several Artemis III candidate landing sites\cite{NASA_CandidateRegions2023} by ray-tracing solar illumination to updated LOLA-based topography maps\cite{BARKER2021105119} with a pixel scale of 5 m/pix. Artemis III is planned to be the first time humans will explore the lunar south pole, as part of NASA's Artemis campaign to re-establish a human presence on the Moon in preparation for future human exploration of Mars. Finally, we briefly discuss the impact of sub-meter scale topographic features due to both anthropogenic and non-anthropogenic sources at relevant polar exploration sites. We find that both lunar poles likely possess significant regions of potential microbial survivability - microbes examined may be able to survive in these regions (including in their PSRs) in a cryptobiotic state, though we do not believe that growth is likely to occur unless habitable conditions are met (see discussion later in study). This finding that has significant implications for both science and exploration requirements and that updates prevailing views that have been influential on policy\cite{national2020report}.


\section*{Results}

\subsection*{Survivability Bounds of Key Relevant Microbes}

Selection of specific microorganisms for our analysis is based upon a number of different factors that reflect important considerations regarding both their ability to survive on the surface at the lunar south pole and likelihood of their delivery to the surface. Some of these particular microorganisms have demonstrated survival in space\cite{doi:10.1089/ast.2017.1751, doi:10.1089/ast.2018.1913, doi:10.1089/ast.2020.2426, OLSSONFRANCIS20101, doi:10.1128/msystems.00195-19}, or have shown they may proliferate in a potential surface payload cabin. Another key factor is how cosmopolitan (common and widespread) a particular microorganism may be. We note that our choices are a small subsample of the diverse range of microbes that may be transferred to regions of exploration, and that they may not include relatively common microbes that are also highly adapted to survival on the lunar poles.

Another key distinction in our analysis is our focus on how survivable a particular niche may be versus how habitable it may be - the threshold for growth is higher than that of survival\cite{Ratliff2023.11.06.565904}. Survival implies that the cell is at least in a dormant state, behaving as a persister cell, is viable but not may not be culturable or has formed a spore\cite{pinto2015thirty}. Cells in any of these survival states can return to more active metabolism when conditions change. We also note that cells may be dead but still persist in the environment because they persist in the environment as potential sources of contamination\cite{lennon2011microbial}. Growth requires active metabolism and potential reproduction. Reproduction can be observed by cultivation while active metabolism can be observed as respiration, enzymatic activity, translation, transcription or changes in membrane potential. Moribund or dead cells will not perform any of these functions and likely will have damaged intracellular biomolecules and phosphorylated membrane lipids. Stains can be used to differentiate cells in this variety of metabolic states\cite{schichnes2006fluorescent, emerson2017schrodinger}.

However, even survival is highly salient to exploration and science - such survival would offer a novel example of the existence of life on another planetary body in the corresponding local environment, is a pre-condition for potential growth, and would also be critical to take into account when conducting astrobiological studies or assessing the local human health environment. The microbes (\textit{Bacillus}\cite{mbs:/content/journal/micro/10.1099/mic.0.000922}, \textit{Deinococcus}\cite{annurev:/content/journals/10.1146/annurev.micro.51.1.203}, \textit{Staphylococcus}\cite{annurev:/content/journals/10.1146/annurev.mi.34.100180.003015}, \textit{Aspergillus}\cite{bennett2007overview} and \textit{Fusarium}\cite{Fusarium}) selected for this study are extremophiles capable of surviving the conditions expected at some lunar locations and/or are cosmopolitan organisms commonly detected in crewed spacecraft.

The microbes are given in Table 1 and are representatives of the microbiome that will likely accompany human space exploration. Common occurrences and motivation for inclusion are described in the second column. We list maximum growth temperatures for each microbe as opposed to survival temperatures in order to use conservative high temperature bounds for survival for a particular microbe and given that relevant studies more commonly measure temperature limits of growth versus survivability. These microbes may survive at higher temperatures, though temperatures in the lunar poles are generally lower than these values. The final column lists empirically derived UV lethal dose values (for each corresponding life cycle stage of a microbe) that we use as a maximum fluence (time integrated flux) above which there is no survival assumed for the particular microbe. Details and assumptions related to how we obtained or estimated these values, as well as related background, is given in Materials and Methods. We do not include values related to vacuum conditions, energetic particle flux, additional radiation sources such as Galactic Cosmic Rays (GCRs) and Earthshine or low temperatures in the table given that they are likely to have negligible relative impact on survival of the chosen microbes over our timescales of interest (see Materials and Methods for analysis of each of these processes).

Maximum growth temperatures for the 5 microbes span a range of less than 20K (the selected fungi are less tolerant to high temperatures than the bacteria), and as detailed in the next section, are all above modeled maximum summer temperatures in the lunar south pole for nearly the entire region. The estimated UV lethal dose fluence, however, varies to a much greater extent. While bacteria such as \textit{Bacillus} and \textit{Staphylococcus} are inactivated at relatively low values, the selected fungi (\textit{Aspergillus} and \textit{Fusarium}) require fluences for inactivation that are orders of magnitude greater than the bacteria. In microbiology, inactivation refers to a set of environmental factors that permanently render an organism incapable of reproducing\cite{rutala2008guideline, VanImpe2018}. Given that temperatures at the lunar poles are rarely high enough to be biocidal and that low temperatures are used to preserve microbes on Earth, UV radiation appears to be the most effective biocidal factor in those regions. Therefore, we focus on mapping UV fluences in the lunar polar regions to identify regions where microbes may survive. We also exclude the few locations where temperatures exceed thresholds needed to inactivate our target organisms.

\subsection*{UV and temperature environments at the lunar south pole}

With the survivability limits for selected microbes from Table 1, based on temperature and UV dose, we can filter for potentially survivable regions of the lunar poles at different spatial scales. A number of data sets were combined to produce regional survivability maps for the north and south pole of the Moon. The regional polar maps \cite{2010SSRv..150...81R} that are used to approximate UV fluence calculate averaged instantaneous UV flux over a lunar day using fractional solar visibility based on LOLA topography\cite{MAZARICO20111066} and an analytical formulation\cite{KOPP2023250} of illumination for flat surfaces based on orbital and latitudinal properties. Temperature data for our regional maps are based on the maximum Summer temperature from LRO Diviner data\cite{https://doi.org/10.1029/2019JE006028}. The top two maps in Figure 1 show regional UV fluence polewards of 85 degrees for the north pole and south pole of the Moon. These maps provide the integrated UV light fluence over 1 Earth day (24 hours) considering direct illumination, based on the average instantaneous flux at the surface. They are meant as a qualitative overview of potential survivability, since their incorporation of topography is at coarser spatial resolution (see Materials and Methods for details). While fluences are similar within an order of magnitude to more accurate high resolution maps (Figure 2C-H), the implicit assumption of flat terrain with regards to local topography in the fractional illumination calculations results in underestimates in more poorly illuminated regions and overestimates in highly illuminated regions. Nevertheless, these maps provide important guidance for potential survivability of different microbes over relevant timelines and can inform selection of smaller areas for more accurate modeling.

Using threshold criteria of a maximum summer temperature under the maximum growth temperature and an integrated average instantaneous UV fluence that does not exceed lethal dose values based on Table 1, the bottom two maps in Figure 1 display an estimate of regions where the microbes of interest likely could survive for at minimum 1 day (24 hours). The top panels of Figure 1 show the significant variability in integrated UV fluence over 1 day near the poles - which spans from roughly 1 MJ/m² in the most highly illuminated regions to no direct UV illumination within PSRs. Fluence also generally decreases towards the pole due to lower Sun elevation angles. Since temperature is highly correlated with UV flux, survivability to temperature broadly follows the trend related to the lower fluence. Modeled conditions in Figure 1 show that \textit{Aspergillus} exhibits the greatest potential for survivability in polar regions with respect to the criteria we identify to be most important, likely due to repair mechanisms that lend exceptional resilience to UV fluence. There are additional regions of survivability for other combinations of the microbes, but they are significantly smaller and generally proximate to PSRs - some of these regions are shown in the magnified maps in the middle of the image, with illustrative sub-regions chosen from the respective pole. These maps are simplified, but show general trends of survivability that provide guidance for selecting specific regions where survivability is modeled more accurately by considering higher spatial resolution properties - details behind both sets of maps are given in the Materials and Methods. Each of the regional maps indicates there are potential regions of survivability in the context of joint ultraviolet flux and high temperature limits for all the microbial species in this study at the north and south pole of the Moon. As we note in the discussion, a more thorough survivability analysis would incorporate higher spatial resolution data and would require additional study of multiple additional and joint factors that may be relevant to survivability through careful experiments and modeling - this study examines survivability in the context of currently understood key factors for survival on the lunar surface exposed to open space.

While regional maps provide a large-scale overview of microbial survivability, the coarser resolution at which they model topography driven shadowing may miss out on potential survivability at finer spatial scales. We thus model varying illumination conditions at sites relevant to Artemis exploration by ray-tracing on recent high resolution LOLA-based topography\cite{BARKER2021105119} (details given in Materials and Methods). We compare seasonal UV fluences over timescales of interest masked with maximum surface temperature values (derived from lower resolution Diviner maps at 240 m/pix) to microbial thresholds from Table 1 to produce survivability maps similar to but more detailed than the regional polar maps shown in Figure 1. These maps are shown in Figure 2 and display the number of surviving microbes at selected Artemis III Candidate Landing Regions after one Earth day in August 2026 (i.e., close to the Sun equinox at the Lunar south pole) and the survivability of \textit{Aspergillus} (the most resilient of the reference microbes from Table 1) over time-scales ranging from a few hours to at least 7 Earth days, for reference. Critically, for these maps we only model direct illumination, since scattered light (particularly in non-PSR regions) has a negligible effect on total survivability timescales and extents (see Materials and Methods for more detail). We do, however, include modeling of scattered light for the case study of a PSR in the De Gerlache region, as it is likely to control survivability timescales. In the Materials and Methods section, we also analyze additional potential sources of UV flux, such as all-sky Lyman Alpha flux and Earthshine, and find that they are likely to have a minimal contribution to survivability limits given the timescales we are examining.

Figure 2A shows the maximum summer temperature close to the Lunar south pole, which is used in all maps as a limit on survivability. Below this map is a table which shows key survival adaptations or factors related to the 5 microbes of interest. To the right of these images are first the maps depicting survivability of the different microbes for 1 day, and then maps showing the average time needed to accumulate a lethal dose of UV flux for \textit{Aspergillus}. These maps differ from figure 1 in that we model illumination by explicitly using ray-tracing on recent high resolution LOLA-based topography. Both selected areas - those in the Nobile Rim and Connecting Ridge region (with De Gerlache listed in table 1, contain significant places where microbes may be able to survive. Of note is that these potentially survivable areas are not just limited to those adjacent to larger PSRs, and in many cases are nearby areas that receive high levels of illumination. \textit{Aspergillus} is the organism that may be able to survive in the largest number of places given our analysis. It may be able to survive in 2-9\% of the mapped (non-PSR) surface area in the summer and 15-30\% of the mapped area in the winter. PSRs are not probed by direct illumination models and so are not included in these calculations even though they receive less radiation and experience lower maximum temperatures than other regions. Fractional areas of survival are roughly 4-7 times lower for \textit{Fusarium}, 5-9 times lower for \textit{Deinococcus}, and 11-30 times lower for \textit{Staphylococcus} and \textit{Bacillus}. The Nobile Rim 2 region contains the highest fraction of potentially survivable areas for the three hardiest microbes during summer and the Connecting Ridge region contains the highest area during winter for all of the microbes. Maximum survivable area and order of total area of survivability varies between all three regions depending on the microbe, for autumn and spring, though Connecting Ridge again contains the largest survivable area for the three hardiest microbes. The De Gerlache Rim region shows the lowest survivable area during summer while Nobile Rim 2 has the lowest survivable area during winter - unsurprisingly survival rates peak in winter and are at a minimum in summer. Critically, all three regions have roughly 3\% survivable area for \textit{Aspergillus} for time scales of at least 7 days, indicating longer term potential survivability (how long merits additional study) of some microbes across these regions of the south pole.

Fluxes within PSRs are set to 0 and masked out (\textit{i.e.}, black) in regional and high-resolution maps presented in Figures 1-2 since we only model direct sunlight for those maps. These regions are the most favorable to survival due to lack of direct illumination, and are likely to have their limits set by scattered light, which is modeled in Figure 3. Total fractional areas of potential survival using our high resolution maps, including separate estimates of PSR areas, are given in Table S1 in Materials and Methods. Since PSRs will be more heavily impacted by scattered sunlight from neighboring illuminated surfaces - as an example, Figure 3 shows survivability within De Gerlache's PSR (88.5°S, 87.1°W) when modeling scattered fluxes using established methods\cite{POTTER2023100130, stefano_bertone_2024_14018762}. We assume surface albedo at UV wavelengths to be the same as for visible light. We show the impact of scattered UV light in limiting survivability areas for multiple microbes, and the estimated life-time for the most resilient species in our set, \textit{Aspergillus}. The impact is mostly relevant within PSRs, as scattered flux represents $\sim1\%$ of the direct solar flux in illuminated areas. Fractional values of survival in the region using this methodology are also given in Table S1 in Materials and Methods. Including scattered light does reduce the total fractional area of potential survival for microbes and total time of potential survival within the PSR, but the fractional area of potential survivability is still much larger than in illuminated regions, as expected. All 5 microbes examined can survive in some regions of the PSR for periods longer than a week, which is broadly similar to the findings of other work \cite{moores2025microbial} that examined survival of \textit{Bacillus} in two PSRs for a specific season and comports with earlier work \cite{glavin2010situ, graham2024survivable} that noted potential survival of microbes in PSRs. Our results are consistent with the assertion that lunar PSRs are, "one of the least biocidal environments in the solar system," \cite{moores2025microbial}. However, differences between these survival times, variations across different PSRs and consideration of other processes that may operate on relevant timescales all motivate the need for higher resolution spatial and temporal studies of these PSRs. Despite that, while our focus is on the general question of potential survivability, studies looking to analyze a more localized region in higher detail should incorporate scattered light, despite the significant additional computational expense.

Analysis of potential survivable niches in the context of UV fluence is relevant at even higher spatial resolution scales - recent Chang'e 3/4/5 mission results show average depth to diameter ratios of smaller craters (d/D of $\sim$ 0.07 for 1-10+ meter craters\cite{2022Icar..37814943B} and d/D of $\sim$ 0.055 for $\lessapprox$ 1 meter craters\cite{2023JGRE..12807703L}) in relatively high latitude regions that suggest additional small scale survivable niches are likely to exist in the polar regions. These two values of d/D would result in permanently shadowed craters on flat terrain polewards of 87°. Importantly, planned human exploration activity is likely to also create similar small scale survivable features - while drill cores and large scoops can create shadowed regions at more equatorial latitudes, even boot prints, rover wheels and small scoops are likely to leave shadowed depressions on flat surfaces at polar latitudes relevant to exploration\cite{Saxena_Steinmetz_2024}.


\section*{Discussion}

\subsection*{Implications for lunar exploration}

The unique properties of the lunar polar regions (both the south and north pole) with respect to exposure to sunlight suggest that there are likely widespread survivable niches for different microbial life in those regions. Vectors for delivery of microbes relevant to those examined in this study are likely to be persistent and frequent due to upcoming exploration (especially since Artemis III candidate landing sites are also candidates for other missions and a base camp) - venting from airlocks\cite{WilliamsISSAirlockVenting, griffin2008lunar, nasa2018office, https://doi.org/10.1029/2023EA003004} and spacesuits that may release organics are also likely to release life, though filtration techniques may mitigate this somewhat. While some of the examined microbes do exhibit survival in spite of clean room contamination control strategies, the presence of humans during exploration inherently increases the number and diversity of microbes that may be transferred since humans themselves are vectors for delivery, given that they carry millions of microbes\cite{scharschmidt2013lives, byrd2018human} on their skin and even more in their bodies. We note that currently there are no bioburden requirements for spacecraft bound for the lunar surface (see \cite{20219}). Best practices to support science efforts motivate re-evaluation of current contamination knowledge-driven planetary protection requirements for Category II permanently shadowed regions \cite{crawford2022managing}. Our findings show that for many investigations it would be prudent to use more stringent contamination mitigation procedures \cite{2024BAAS...56a.009K}. While we view transfer during exploration as the most likely delivery mechanism of life to the polar regions, but it is important to note other means of transfer of material to the lunar poles. For example, meteoritic transfer from outside the Earth-Moon system\cite{Brinton1996, 2015AsBio..15..154M} may be able to deliver organics to the lunar surface while transfer of both meteorites\cite{2002Icar..160..183A, 2010EM&P..107...43A} and molecules\cite{2017NatAs...1E..26T, 2020GeoRL..4786208W, 2025ComEE...6.1001P} from the Earth may be able to deliver organics and/or life, though in all cases survivability will be controlled by the dynamics of the delivery.

Since survival of microbial life may be plausible given our analysis related to the currently understood key factors, a natural question is whether there may exist habitable conditions that enable growth for survival candidates. We focused on survival as a less stringent precursor to growth partially due to our belief that the lack of a stable, dense atmosphere is likely to preclude a key growth factor, \textit{i.e.}, the existence of liquid bioavailable water at the surface. While periods in the geological past\cite{doi:10.1089/ast.2018.1844} may have been habitable windows due to the brief existence of collisional atmospheres\cite{NEEDHAM2017175, 2017E&PSL.474..198S}, the only potential present day atmospheres\cite{PREM2015148, KILLEN2024} are likely to be highly transient and localized, though we note that anthropogenically generated atmospheres may be concurrent with delivery of microbial life. Concurrent potential delivery of organics that can serve as nutrients at the same time as delivery of microbial life\cite{https://doi.org/10.1029/2023EA003004} may be a growth enhancing factor, as may be modification of the surface to create trapped water that may episodically melt [see previous work on potential short lived subsurface lunar water\cite{STOPAR2018157} and fluidization studies on comets\cite{BELTON2009280, SUTTLE2020113956}], but both those mechanisms require additional study. Importantly, modification of the lunar environment due to human exploration that will inherently occur episodically but concurrently with potential transfer of life may result in novel processes relevant to these questions.

The plausibility of survival of microbial life in these regions is predicated on existing experimental work and modeling derived from existing remote sensing data. Additional study is needed in numerous areas for more refined analysis of potential survivable niches for microbial life - these additional studies should include experimental work that examines key additional properties of the lunar surface that may inhibit survival, the interactive effects of such properties, broader and higher spatial and temporal resolution studies of survival, and a more thorough and focused examination of work related to microbial survivability that has previously been carried out in a more general way. For example, while most regions do not likely have temperatures high enough for the biocidal interactive effects of the combination of simultaneous high temperatures (near or above growth limiting temperatures) and ultraviolet radiation \cite{doi:10.1089/ast.2018.1952} to become critical, this interaction may be true in some small pockets and additional experimental work for lunar analogue environments would be illuminating. Understanding the number, diversity, and communities of microbes that may be released is another key area of future work - even dead microbes on the lunar surface may complicate astrobiological and other studies. Experimental UV light lethal dose studies that extend to lower survival fractions in lunar analogue environments would also provide more accurate estimates. From the remote sensing perspective, higher resolution Digital Elevation Models (DEMs) of the lunar surface down to sub-cm scales would enable more accurate modeling of key factors relevant to potential survivability - this is not limited to the poles, as other natural features (\textit{e.g.} pits and dikes) can generate niches at lower latitudes.

Understanding complicating effects of geology and other geophysical processes, surface modification due to human exploration and potential effectiveness of different mitigation techniques\cite{lombini2023solar} are all also areas of potential future work. For example, models of impact gardening predict an average of equilibrium excavation and burial that would impact the top micron of the regolith on a timescale of roughly 10 years and the top centimeter on a timescale between 50,000-100,000 years\cite{2020JGRE..12506172C, costello2021secondary} -- while the time to impact microbes at these depths is longer than timescales of interest in this paper, such a process should be considered for longer term survivability studies. Gardening of the regolith by human exploration is likely to impact a range of different depths depending on direct or indirect contact with the surface, and consequently emplacement of microbes may be at varying depths. In fact, many of the processes that we have been able to ignore as negligible over this study's timescales of interest may have non-negligble impact over longer time periods and should be considered for relevant survivability studies (see Materials and Methods).

We note that studying such questions, and related ones such as those concerned with locations that may be survivable as well as limiting transfer, have applicability beyond the Moon, and are an opportunity for science, not necessarily merely a hindrance. Indeed, a survivable niche may not necessarily be a place to avoid during exploration, but instead one to target for sample acquisition (potentially repeatedly) using a well thought out sampling strategy. A number of different airless bodies such as Mercury, Ceres, Asteroids, Comets, and Exoplanets that may possess similar properties could also possess potential survivable niches for microbial life. Finally, understanding how these survivable niches may exist, how life may be transferred to them, and how human exploration may leave its mark all will serve as a key testbed for future potential human exploration of Mars, which likely possesses far more habitable environments.


\section*{Materials and Methods}

The following sections contain information on the methodology and assumptions that contributed to the findings in the main body of the paper.

\subsection*{Microbe properties: assumptions and derivations}

This section details assumptions and underlying derivations and data that was used to derive values in Table 1, which were then compared to lunar surface properties. As stated in the main text, the key factors examined in this study with respect to microbial survival were UV radiation and high temperatures. Vacuum conditions, energetic particle flux and low temperatures were not included due to evidence from literature they were not dominant or significant factors in inhibiting survival of the chosen microbes over the timescales examined for this study. We discuss why this assumption was made and the evidence in the literature that underpinned it in the following text.

The surface of the Moon is exposed to a surface bounded exosphere that is close to a vacuum. Microbes can survive exposure to the vacuum of space (\cite{doi:10.1128/MMBR.00016-09, 10.3389/fmicb.2019.00780} - see additional references in section 1 which document survival after exposure to open space) without experiencing more than a 10-fold decrease in the number of viable cells for timescales significantly longer than those relevant to this study. In fact, the combination of cold temperatures and exposure to vacuum (freeze-drying or lyophilization) is a common method used to preserve microorganisms for long term storage on earth\cite{prakash2013practice}. This specifically includes studies that examined survival of exposure to vacuum for some of the microbes we examine, \textit{e.g.}, \textit{Bacillus}\cite{HORNECK199441,doi:10.1089/ast.2018.1952} (in the latter study, a 1 log reduction took roughly 1000 days), \textit{Deinococcus}\cite{doi:10.1089/ast.2018.1913}, and \textit{Aspergillus}\cite{DOSE1995119, SARANTOPOULOU201163} (though additional lower pressure studies have not yet been conducted). Given the timescales of exploration we are interested in or even longer periods such as the lunar synodic period, vacuum effects are unlikely to significantly inhibit survival.

Similarly, over the timescales of interest (1 day to 1 week), we assume energetic particle radiation has a limited impact on the survivability of the microbes we are examining due to evidence in the literature regarding the weaker effect of such a mechanism. A broad examination of the effect of ionizing radiation on fungi found even 1 log reductions of included and relevant fungi and bacteria required nearly kGy levels of ionizing radiation\cite{dadachova2008ionizing}. In-situ observations of energetic particle flux on the lunar surface\cite{zhang2020first} and in orbit near the Moon\cite{george2024space} indicate that on the timescales of interest, total energetic particle dosage is likely to be many orders of magnitude below this. In fact, recent research\cite{2024LPICo3040.1513P} into relevant energetic particle radiation dosages at the lunar south pole indicates radiation dosages roughly at the $10^{-3}$ Gy level over a 30 day period during quiescent solar activity periods and $\sim$40 Gy for the most energetic Coronal Mass Ejection (CME) events recorded (based on the highly energetic 1989 CME event). Recent research indicates a significantly less than 1 log reduction in wild type strains of \textit{Aspergillus} given exposure to radiation dosages similar to the CME event using different types of energetic particle radiation\cite{10.3389/fmicb.2020.00560}. For exposure to radiation dosages relevant to quiescent sun exposure, experiments suggest there is barely any reduction in total population. The effect is observed to be similar for \textit{Bacillus subtilis} based upon experiments conducted that examined survivability given exposure to varying radiation dosages\cite{MOELLER2010783, doi:10.1128/jb.00018-07}. Those studies also found a significantly less than 1 log reduction at the relevant radiation dosages. Total energetic particle flux including solar energetic particles result in lethal dose timescales of 100s-10000s of years, depending on assumptions related to solar energetic particle flux (lower times assume constant powerful CME associated fluxes). GCR fluxes are even lower than these solar sourced energetic particle flux radiation dosages - based on the most recent in-situ measurements from LRO CRaTER\cite{looper2013radiation} and RADOM results on Chandrayaan 1\cite{dachev2011overview} (convolving fluxes with particle cross sections gives destruction timescales of more than 1 million years). Similarly, 'Earth Wind' particles that impact the Moon from inside the Earth's magnetosphere are much smaller in magnitude than solar energetic particles (with energy fluxes several orders of magnitude smaller\cite{wang2021earth}, including during sub-storms\cite{harnett2013substorm}) and also consequently similarly have relatively minimal effect on survivability of the relevant microbes on the timescales we examine.

Low temperature survival of microbes for lunar surface temperatures relevant to this study is supported by a substantial body of literature (see Section 1 for references of survival of microbes exposed to open space). Survival of microbes during exposure to such temperatures during space based experiments on survivability are cited in the introduction and provide direct evidence of the ability of relevant microbes to survive temperatures similar to those on the lunar poles. Dehydration techniques that preserve survivability at similar temperatures are also commonplace for relevant microbes\cite{prakash2013practice}. Finally, a number of experimental studies have examined this survivability and found minimal to no effect on (and in some cases, increased) survivability of microbes at relevant lower temperatures (studies relevant to \textit{Bacillus} and \textit{Aspergillus} cited as an example)\cite{1985Natur.316..403W, doi:10.1089/ast.2019.2214,dose1996response, SARANTOPOULOU201163}.

UV lethal dose values in Table 1 are calculated based upon empirical studies referenced in the same table. For a combination of simplicity and due to lack of empirical studies that extend down to 6 log reductions, we use single stage decay models based upon empirical data in order to determine the fluence required for a lethal dose (D99.9999, 6 log reduction in the number of viable organisms \cite{kowalski2010ultraviolet, doi:10.1089/ast.2018.1952}) which the medical community defines as functionally sterile\cite{internationalorganizationforstandardization} - this is approximately equivalent to a Sterilization Assurance Level (SAL)\cite{Craven2021} of $10^{-6}$. We use the empirical fluence value at the lowest survival rate given in order to determine the D99.9999 value so that we minimize the extent of extrapolation, and use multiple points along the curve to fit the exponential decay. In the cases of \textit{Aspergillus} (wild type) and \textit{Deinococcus}, we use the fluence at 3 log reductions in order to extrapolate a lethal dose. For \textit{Fusarium}, \textit{Bacillus}, and \textit{Staphylococcus} we use D90 derived values to extrapolate lethal dose fluence.

We only choose inactivation studies that are conducted in air or on surfaces, since inactivation fluence can vary significantly in studies conducted in a liquid medium due to attenuation effects. We also choose studies where experiments are conducted in dry air, given the potential influence of humidity on survival\cite{kowalski2010ultraviolet}. In general, our choices to calculate UV lethal doses lead to underestimates of the likelihood of survival for the microbe, \textit{e.g.}, we will not be considering two stage decay and shoulders in the decay rate curves - both features decreasing the susceptibility to UV radiation. Additionally, we do not consider photoreactivation and photorecovery\cite{doi:10.1128/aem.49.4.975-980.1985}, as well as other repair mechanisms, for simplicity of analysis. Considering some of the timescales for reduction and the potential effectiveness of photorepair at low UV flux levels may be a factor increasing survivability and should be considered in future work. We choose microbe states - spores or vegetative states, based upon the hardiest state for a particular microbe that was likely to exist in a human inhabited space and could consequently be transferred to the surface.

\subsection*{UV and temperature calculation methodology}

UV fluence and maximum temperatures for the polar region maps are derived from pre-existing illumination and temperature maps for the south and north pole. For these regional maps, maximum Summer temperatures from previous work\cite{https://doi.org/10.1029/2019JE006028} (with a ground resolution of 240 m/pix) are compared to maximum growth temperature thresholds in order to ascertain survivability. The polar region maps of the UV fluence for the lunar south and north pole are created using the following methodology.

In these polar region maps, fractional illumination over a lunar synodic period for the lunar poles is given in previous studies\cite{MAZARICO20111066} at with a resolution of 60 m/pix based on modeling of illumination conditions over a full 18.6 year lunar precession cycle. We combine the analytically (see Eq. 6 from \cite{KOPP2023250}) integrated fractional illumination received at a particular location with the latitude dependent UV flux in order to produce an estimate of the average instantaneous flux at each surface location. The analytical formulation\cite{KOPP2023250} of illumination is based on orbital and latitudinal properties at a particular point on the surface and we calculate the fluence in 0.05 degree latitudinal increments ($\sim$1.5 km). The inclination of the Moon's rotational axis relative to the Sun is set to 1.5424 degrees, with calculations taken for `summer' periods in order to remain conservative regarding survivability. Bounds in Eq. (6) are based upon longitudinal terminator values given variations in the rotational axis relative to the Sun, and time of the year. While this does apply to the Moon, daily illumination values from previous work\cite{MAZARICO20111066} include this factor and have more detail due to incorporation of topography and averaging over a lunar nutation cycle. As a result bounds are taken as -90 to 90 degrees longitude to have the normalized flux when incorporating the higher detail fractional illumination values. These calculations also assume flat surfaces within the discrete element of calculation, which is relaxed in our later higher resolution modeling. Once we obtained an integrated value, we then scale the total value of the fluence by the total fractional Sun visibility, which assumes equal parts day and night based on the longitudinal bounds.

The total fluence is then divided by the total time in the synodic period to obtain the averaged instantaneous flux - this is an approximate estimate of flux that can be used for daytime estimates of total flux over a period. Since this includes periods where a region is not illuminated, it will be an underestimate for integrated flux during times around noon for a location and will obviously be an overestimate during nighttime, dawn and dusk periods. Despite these divergences at different times, this approximation is useful because it acts as a guidepost to making estimates of total UV fluence for a location and because accurate fluxes at higher spatial resolution are strongly influenced by local topographic effects, which are only calculated for our high resolution analysis of sites. Operational, predictive estimates should be made using high resolution models.

Regional and higher resolution maps in Figures 1 and 2, respectively, calculate direct illumination UV flux from previous studies\cite{https://doi.org/10.1029/2008GL036373} for wavelengths shortwards of 320 nm based on comparisons to previous work and reduced inactivation at longer wavelengths\cite{doi:10.1089/ast.2018.1952, kowalski2010ultraviolet}. Neglecting reflected light in these maps as a first order approximation is a fairly accurate estimate given low albedos for most common minerals and surfaces at the lunar surface (plagioclase being a notable exception)\cite{CLOUTIS2008321, IZAWA2014157, WAGNER198714} and the fact that scattered light is at most $\sim$1\% of the total in illuminated regions\cite{martin2024imaging}. Indeed, time dependent higher spatial resolution observations of different regions of the Moon (including those near PSRs) show that direct illumination is anywhere from 3-5 order of magnitude greater than scattered light (5 orders in figure 2 of\cite{9755998} and roughly 3 orders in figure 8 of\cite{MAZARICO20183214}). However, as mentioned in the interpretation of Figure 3, incorporating scattered light is necessary when probing PSRs within higher resolution maps meant to inform exploration.

High resolution maps in Figures 2 and 3 model instantaneous surface UV fluxes for wavelengths shortwards of 320 nm also based on Whole Heliosphere Interval (WHI) data\cite{https://doi.org/10.1029/2008GL036373} and incorporate a finite angular size for the Sun. Earthshine and all-sky stellar sources of light are not included due to their relatively low flux compared to solar sources. While Earthshine can be comparable to scattered solar illumination in the most favorable circumstances\cite{GLENAR2019841} and may be worth including in higher detail maps in specific cases, all-sky stellar sources are roughly 3 orders of magnitude weaker than scattered solar illumination\cite{KLOOS2021432} and would consequently operate on much longer timescales than relevant to this study. Illumination modeling is carried out by ray-tracing on 5 m/pixel LOLA-based topography with the ShadowSpy tool\cite{stefano_bertone_2024_14018762}, which leverages state-of-art ray-tracing libraries\cite{Woop2024RenderKit,cgal:f-i-24b} and planetary ephemerides from the NAIF/Spice toolkit\cite{ACTON199665,ACTON20189}.

\subsection*{Methodology of survivability based on maps}

The global datasets mentioned previously were gathered for combined analysis into a Geographical Information System (GIS), using ESRI ArcMAP software (ArcGIS Desktop 10.8.2), for the regional north and south polar maps. The base maps used for Figure 1 for the polar regions are mosaics from the Lunar Reconnaissance Orbiter Camera (LROC) Wide Angle Camera (WAC) allowing a pixelated view of the 100 m/pix data. Regional maps of survivability for the north pole and south pole mask the joint UV illumination and temperature maps for each microbe using the survivability criteria described previously (Table 1) integrated over relevant timescales. Importantly, these regional maps provide a \textit{qualitative} estimate of survivability due to the assumptions in the calculation of the UV fluence. The general effect of these assumptions is to provide an overestimate of fluence in regions that receive less illumination and an underestimate in well illuminated regions - these variations are typically less than a factor of 3 but peak at an underestimate of roughly a factor of 8 in the most well illuminated regions and an overestimate by a factor of 12 in the most poorly illuminated non-PSR regions. In rough descending order of importance, differences are due to: 1) The flat terrain assumption used when applying equation 6 in previous work\cite{KOPP2023250}. Fractional illumination values are often driven by significant local topography which results in dramatic deviations of the local elevation angle of the Sun from the latitude-based flat terrain assumption. Due to the general low elevation angle for flat terrain, these differences can result in multiple factors divergences in total flux. 2) Applying a fractional illumination scalar to the regional maps involves scaling a half day/half night illumination profile by the fractional value based on averaging over a full nutation cycle. However, elevation angle, and consequently flux profiles, are dependent on local elevation angle and fractional illumination values and consequently may diverge from this scaling. 3) Longitudinal bounds in the integration of equation 6 in previous work\cite{KOPP2023250} are taken to be -90 and 90 degrees, but these will be slightly different given the orbital and obliquity values and time of season. While divergences in fluence versus higher resolution maps may be large for specific places, these regional maps provide a good general guide for areas that may offer enhanced or diminished survivability - high resolution maps should be used for accurate analysis of specific sites.

Based on the same dataset as Figures 2 and 3, Table S1 presents the fraction of surface within each site where different microbes are likely to survive for 24 hours after deposition, based on the cumulative UV flux received at the surface. As expected, UV flux depends on seasonality (which determines the elevation of the Sun above the local horizon) so that the extent of survivable areas is larger in the "local" Winter than in Summer. Since the cumulative UV flux is based on seasonal average instantaneous UV fluxes, the reported values might still change on a daily basis. The Permanently Shadowed Regions (PSR) column indicates the fraction of each site which never receives direct sunlight during each season: while bacteria are likely to have higher survivability rates in these areas, more detailed analyses are needed to assess the impact of UV light scattered from neighboring (illuminated) areas. As an example, we performed such analysis at the PSR within De Gerlache polar crater, by considering scattered UV light (we assume an albedo of 0.1, a typical value for visible light on the Moon) from a region with a radius of 40 km around the PSR. While scattered UV flux is $\sim$1\% of the direct flux, only \textit{Aspergillus} shows a widespread survivability within the PSR, thus highlighting the need for more detailed modeling. Table S2 shows the portion of each site where \textit{Aspergillus} is expected to survive over time at different target sites of interest. One can note how even after 5-7 Earth days, non negligible portions of the surface are still survivable after deposition in these regions where UV fluxes are relatively limited.


\clearpage 


\bibliography{sn-bibliography} 
\bibliographystyle{sciencemag}



\section*{Acknowledgments}
The authors would like to acknowledge Jay Friedlander for his work and advice in helping produce Figures 1 and 2. P.S. and H.V.G. would like to thank Jason Dworkin and Barbara Cohen for conversations that helped to improve the quality of this work. P.S. and H.V.G. would also like to acknowledge support from the NASA Goddard Science Task Group Program. P.S. and N.M.C. would like to thank Agatha S. and Ayden C.-B. for helping underscore the significance of life at the smallest scales. P.S., S.B., H.V.G., and N.M.C. would like to acknowledge partial support of this work through a grant awarded under 24-PPR24-0003 of the NASA ROSES Planetary Protection Research program. P.S. would like to acknowledge support from the Goddard Space Flight Center (GSFC) Sellers Exoplanet Environments Collaboration (SEEC), which is supported by the NASA Planetary Science Division's Research Program. S.B. acknowledges support by NASA under award number 80GSFC24M0006 and by the ISFM work package Planetary Geodesy at Goddard Space Flight Center. H.V.G. would like to acknowledge support from the CIFAR Earth 4D Subsurface Science Program. Computational resources supporting this work were provided by the NASA Center for Climate Simulation (NCCS) at Goddard Space Flight Center.

\paragraph*{Funding:}
This work was supported by NASA grants 24-PPR24-0003 (P.S., S.B., H.V.G., N.M.C.), 80GSFC24M0006 (S.B.), the NASA Goddard Science Task Group Program (P.S., H.V.G.), and the GSFC Sellers Exoplanet Environments Collaboration (P.S.). Additional support for all authors was also provided by the National Aeronautics and Space Administration (820GSFCS).

\paragraph*{Author contributions:} 
P.S. formulated the concept of the paper, organized the structure, and led interpretation of the implications of the results. P.S. also led writing of the paper, contributed to investigation, execution of methodology, project administration, and guided broader analysis of microbial survivability implications. N.M.C. carried out data curation, formal analysis, and visualization, developed software, and contributed to resources, validation, and writing of the paper. N.M.C. also led the creation of regional UV microbial survival maps, interpreting their implications and contributing to conceptualization, investigation, and methodology. S.B. led high-resolution mapping and analysis of microbial survivability in Artemis Landing Site Candidate Regions, including modeling of both direct sunlight and scattered light. S.B. also contributed to funding acquisition, formal analysis, methodology evaluation, visualization, and writing and editing the paper. H.V.G. contributed to conceptualization, writing of the paper, and investigation, leading candidate microbe selection and curation of microbial properties. H.V.G. also supported methodology evaluation, validation, and visualization efforts. A.B.R. contributed to conceptualization, candidate microbe selection, curation of microbial properties, interpretation of survivability data, and writing of the paper, as well as methodology development. A.N. contributed to conceptualization, writing and review of the paper, validation, and formal analysis. D.E.P. and N.E.P. contributed to conceptualization, writing and review of the paper, validation, analysis of broader results, methodology development, and visualization support.

\paragraph*{Competing interests:}
The authors declare no competing interests.

\paragraph*{Data, Code, \& Materials Availability:}
The authors declare that the data supporting the findings of this study are available within the paper, its supplementary information files, and from maps hosted at the Planetary Geodesy Data Archive (https://pgda.gsfc.nasa.gov/products/78). The code used to produce high resolution survivability maps used in the analysis is publicly available at https://doi.org/10.5281/zenodo.14018762. This study did not generate new materials.



\begin{figure}
\centering
\includegraphics[width=\textwidth]{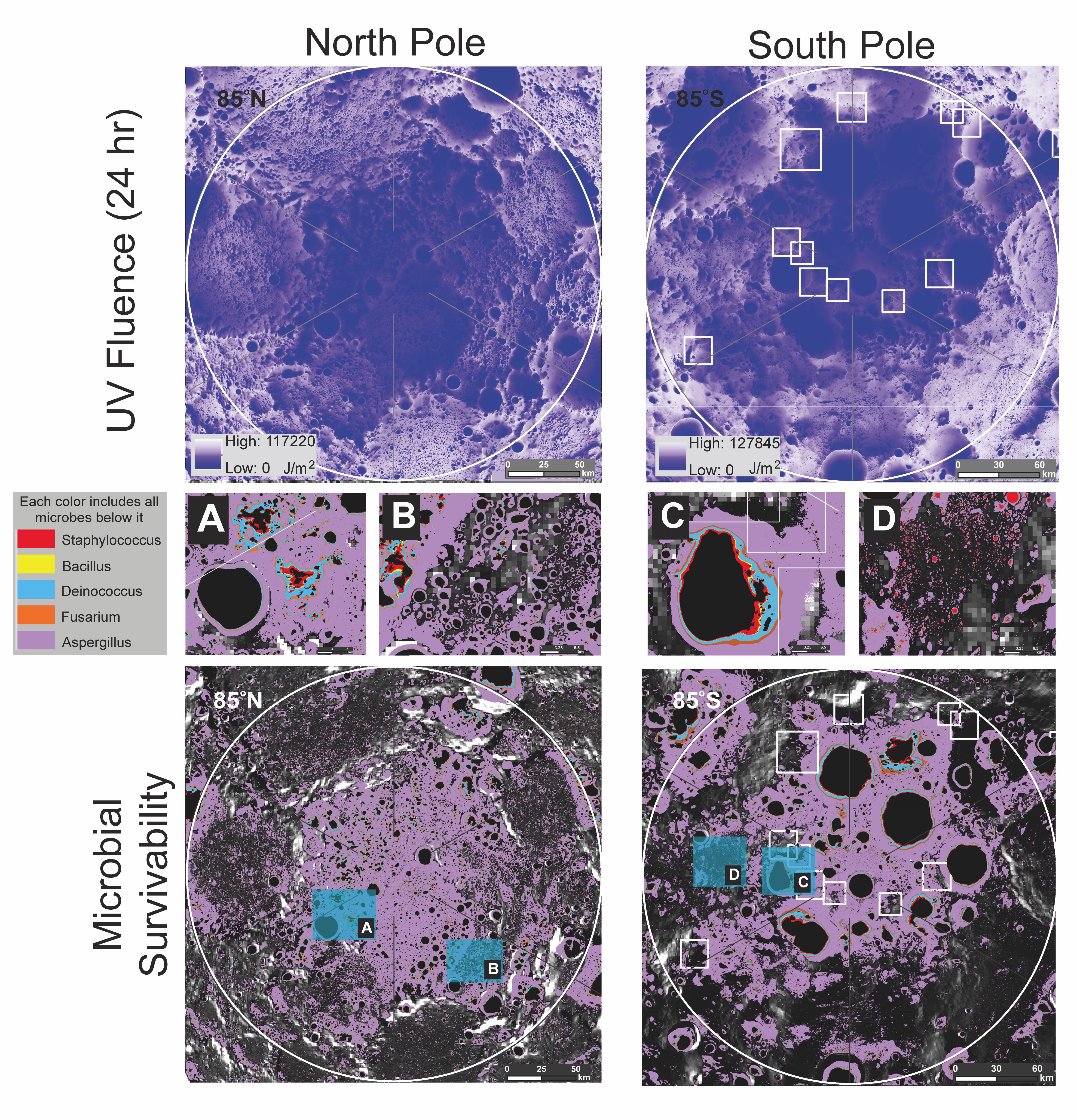}
\caption{\textbf{Microbial survivability in the polar regions of the Moon.} Top panels show the UV fluence over a 24 hour period in the polar regions of the Moon. Bottom panels show areas where each microbe may be able to survive based on corresponding limits related to the direct integrated UV fluence and maximum summer temperature (with PSRs, given in black, as survivable for all microbes under this scenario). \textbf{(A)} North pole enlarged region showing survivability for multiple microbes in the vicinity of larger permanently shadowed regions. \textbf{(B)} North pole enlarged region displaying survivability patterns for multiple microbe species interspersed with areas of lower survivability. \textbf{(C)} South pole enlarged region near the De Gerlache region showing survivability of multiple microbes surrounding the permanently shadowed region. \textbf{(D)} South pole enlarged region showing survivability patterns for either multiple microbe species or only \textit{Aspergillus} interspersed in large area of lower survivability. White squares on the bottom panels show the Artemis III candidate regions. Round circles on each of the large panels show the 85° North or South.}
\label{fig:reg_maps}
\end{figure}

\begin{figure}
\centering
\includegraphics[width=\textwidth]{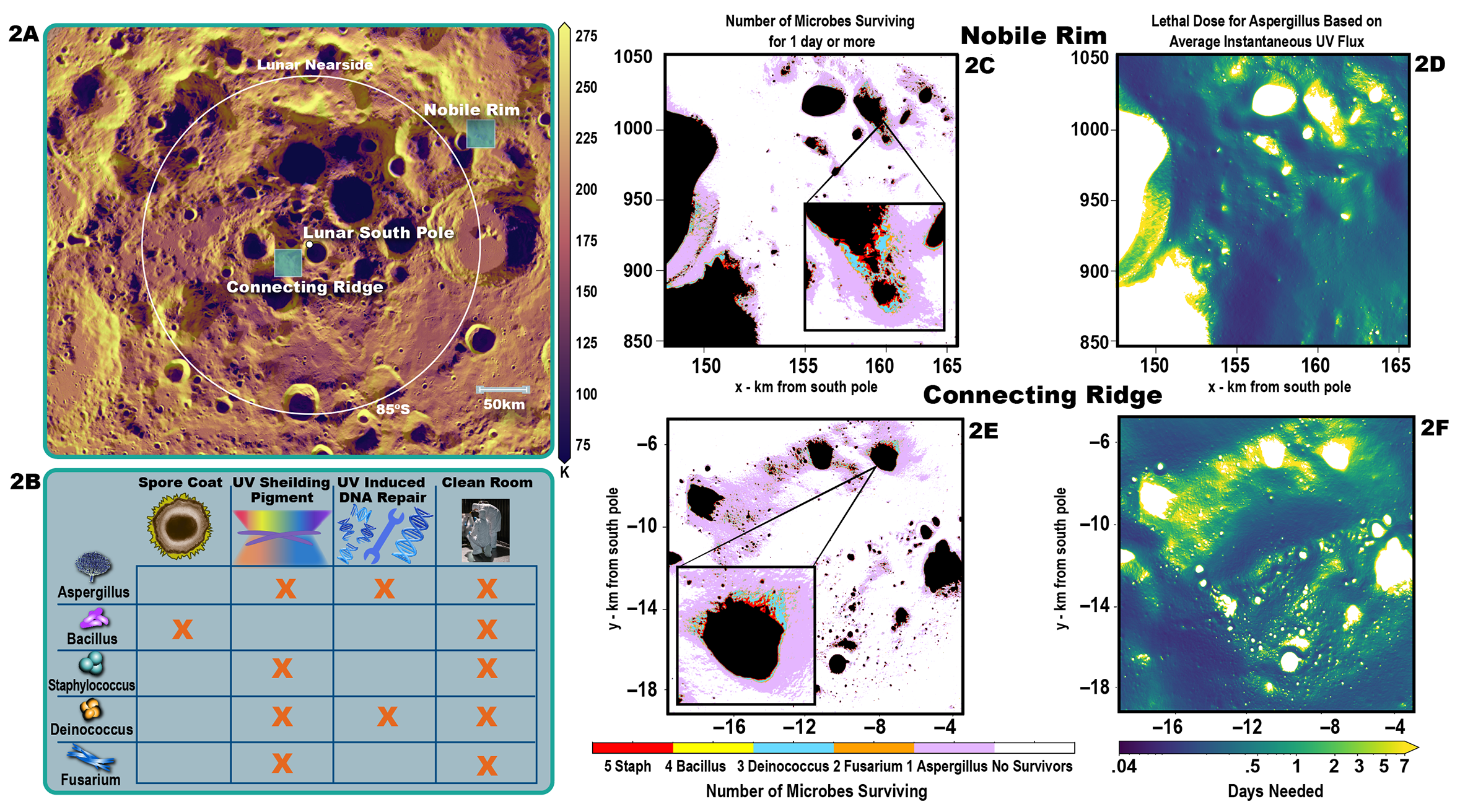}
\caption{\textbf{Microbial survival mechanisms and lunar surface conditions at Artemis III sites.} \textbf{(A)} Maximum summer temperature map showing the thermal environment at the lunar south pole, with cooler temperatures (darker colors) providing more favorable conditions for microbial survival. \textbf{(B)} Survival mechanisms table showing key adaptations of the five examined microbes, including spore formation, UV-protective pigments, and DNA repair capabilities, that enable survival in harsh lunar conditions, as well as clean room prevalence, which makes transfer more plausible. \textbf{(C)} Number of microbial species surviving after one Earth day at Nobile Rim region, showing spatial distribution of survivability with up to 5 species surviving in the most favorable locations. Similar to figure 1, the larger the number includes all the species listed below the color. \textbf{(D)} Similar to \textbf{(C)}, number of microbial species surviving after one Earth day at Connecting Ridge region, demonstrating areas where multiple species can survive simultaneously. \textbf{(E)} \textit{Aspergillus} survivability duration map for Nobile Rim showing time periods (in days) that this most resilient species can survive, with maximum survival times of at least 7 days in optimal locations. \textbf{(F)} Similar to \textbf{(E)}, \textit{Aspergillus} survivability duration map for Connecting Ridge region, indicating extended survival periods with some areas supporting survival for 7+ days.}
\label{fig:lsp_surv}
\end{figure}

\begin{figure}
\centering
\includegraphics[width=\textwidth]{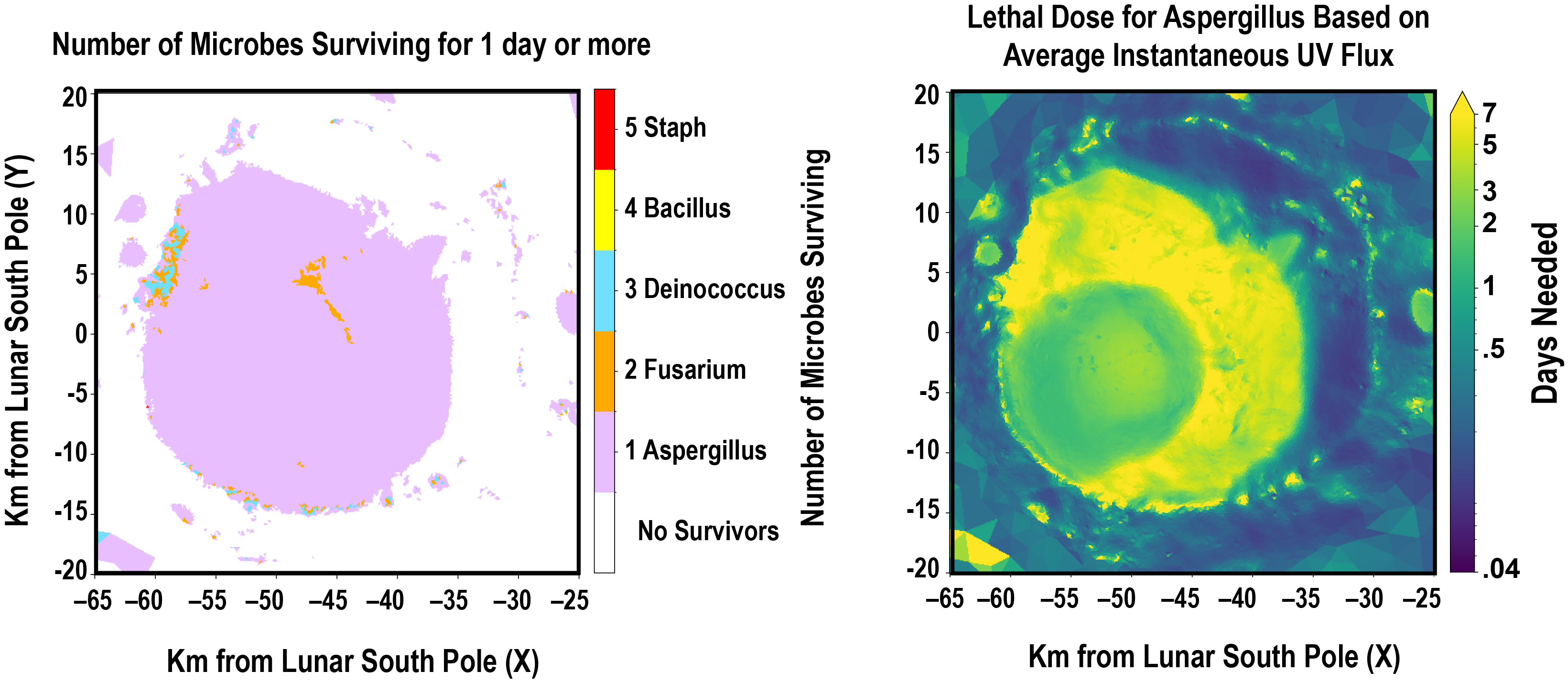}
\caption{\textbf{Impact of scattered UV light on microbial survivability within permanently shadowed regions.} Lunar PSRs never receive direct sunlight because Sun's elevation never exceeds their local horizon. However, flux scattered by nearby topography still reaches them. We model scattered UV flux within De Gerlache's PSR to highlight its impact on survivability. \textbf{(A)} Microbial survivability map showing which species can survive at least one day within the PSR when accounting for scattered UV light, with the same color scheme as in Figure 1 (\textit{Aspergillus} in cyan showing the most extensive survival areas, followed by other species in decreasing order of survivability). \textbf{(B)} \textit{Aspergillus} survival duration map showing survival time in days within the PSR, with the most favorable areas (yellow) supporting survival for the maximum modeled period of at least 7 days, while other areas show progressively shorter survival times (darker colors).}
\label{fig:scat_dg0}
\end{figure}


\begin{table}
\centering
\caption{\textbf{Survivability assessments for microbial candidates.} Lethal UV doses are given based on the environment of the experiment (irradiation conducted in air or on a surface, Relative Humidity (RH) amount if in air) and the status of the organism with respect to whether cell in the experiment is in a vegetative (veg) or spore state. All growth values are for vegetative state.}
\label{tab:microbes}
\resizebox{\linewidth}{!}{%
\setstretch{0.8}
\small
\renewcommand{\arraystretch}{1.3}
\begin{tabular}{|C{2.2cm}|C{4cm}|C{2.8cm}|C{4.2cm}|}
\hline
\normalsize\textbf{\textit{Microbe}} & \normalsize\textbf{\textit{Places Commonly Found/Motivation for Inclusion}} & \normalsize\textbf{\textit{Max Growth Temperature dry (K)}} & \normalsize\textbf{\textit{UV Lethal dose (6 log reduction) in J/m²}} \\
\hline
\textit{Bacillus} (\textit{subtilis}) & Very common human-associated heat-resistant mesophile spore-former & \textbf{328}\cite{mbs:/content/journal/ijsem/10.1099/00207713-39-3-295} & 84 (veg, air/low RH)\cite{nakamura1987sterilization}, \textbf{410 (endospore, surface)}\cite{munakata1975effects} \\
\hline
\textit{Deinococcus} (\textit{radiodurans}) & Radiation resistant, demonstrated survival in exposure to space & \textbf{331}\cite{doi:10.1089/ast.2018.1913} & \textbf{2,280 (veg, surface)}\cite{mbs:/content/journal/micro/10.1099/00221287-129-8-2437} \\
\hline
\textit{Staphylococcus} (\textit{aureus}) & Very common heat-resistant human-associated mesophile & \textbf{324}\cite{medvevdova2010effect} & \textbf{300 (veg, surface)}\cite{hollaender1954radiation} \\
\hline
\textit{Aspergillus} (\textit{niger}) & Radiation resistant, common ISS contaminant & \textbf{315}\cite{10.1046/j.1365-2672.1997.00297.x} & 27,000 (veg, surface)\cite{chick1963ultraviolet}, \textbf{30,560 (spore, surface)}\cite{10.3389/fmicb.2020.00560} \\
\hline
\textit{Fusarium} (spp.) & Common heat-resistant soil fungus and ISS contaminant & \textbf{314}\cite{bennett2012survival} & 6,720 (veg, surface)\cite{chick1963ultraviolet}, \textbf{3,360 (spore, surface)}\cite{chick1963ultraviolet} \\
\hline
\end{tabular}%
}
\end{table}



\clearpage 


\renewcommand{\thefigure}{S\arabic{figure}}
\renewcommand{\thetable}{S\arabic{table}}
\renewcommand{\theequation}{S\arabic{equation}}
\renewcommand{\thepage}{S\arabic{page}}
\setcounter{figure}{0}
\setcounter{table}{0}
\setcounter{equation}{0}
\setcounter{page}{1}


\begin{center}
\section*{Supplementary Materials for\\ \scititle}

Prabal~Saxena$^{\ast}$,
Stefano~Bertone,
Heather~V.~Graham,
Natalie~M.~Curran,
Aaron~B.~Regberg,
Andrew~Needham,
D.E.~(Betsy)~Pugel,
Noah~E.~Petro\\
\small$^\ast$Corresponding author. Email: prabal.saxena@nasa.gov

\vspace{1cm}
\textbf{Table of Contents:}\\
Tables S1 to S2\\
\end{center}

\newpage


\begin{table}
\centering
\caption{\textbf{The percentages of the surface where each microbe can still survive after 24 hours of average UV flux.} The table shows the percentage of each region's surface where different microbes are likely to survive based on lethal doses from Table 1 and total UV flux received. The first column indicates the fraction of a region that is a PSR - this region would receive no direct sunlight and thus in this illumination scenario, all 5 microbes would be able to survive. The following columns indicate the non-PSR regions where the respective microbe would survive according to the above criteria. All of the rows without a * only model direct flux. The * indicates that both direct and scattered fluxes have been modeled and accounted for, which is why the PSR region is set to 0 since it is a recipient of scattered flux.}
\vspace{4pt}
\label{tab:surviveregions}
\begin{tabular}{lcccccc}
\hline
\textbf{Site} & \textbf{PSR Fraction} & \textbf{\textit{Aspergillus}} & \textbf{\textit{Fusarium}} & \textbf{\textit{Deinococcus}} & \textbf{\textit{Staphylococcus}} & \textbf{\textit{B. Subtilis}} \\
\hline
\multicolumn{7}{c}{Spring/Autumn (Survivable Surface, \%)} \\
\hline
Nobile Rim 2 & 15.71 & 12.51 & 3.18 & 2.59 & 1.12 & 0.97 \\
Connecting Ridge & 6.55 & 16.75 & 3.15 & 2.55 & 1.17 & 1.03 \\
De Gerlache Rim & 5.55 & 13.84 & 2.42 & 1.89 & 0.68 & 0.57 \\
De Gerlache PSR* & 0.00 & 40.19 & 1.35 & 0.56 & 0.00 & 0.00 \\
\hline
\multicolumn{7}{c}{Summer (Survivable Surface, \%)} \\
\hline
Nobile Rim 2 & 10.81 & 9.23 & 1.45 & 1.03 & 0.33 & 0.28 \\
Connecting Ridge & 3.55 & 7.11 & 1.06 & 0.86 & 0.40 & 0.35 \\
De Gerlache Rim & 2.01 & 2.75 & 0.47 & 0.39 & 0.18 & 0.16 \\
De Gerlache PSR* & 0.00 & 25.39 & 0.03 & 0.01 & 0.00 & 0.00 \\
\hline
\multicolumn{7}{c}{Winter (Survivable Surface, \%)} \\
\hline
Nobile Rim 2 & 23.18 & 15.43 & 3.04 & 2.37 & 0.91 & 0.79 \\
Connecting Ridge & 19.93 & 30.26 & 5.98 & 4.29 & 1.48 & 1.29 \\
De Gerlache Rim & 42.14 & 20.44 & 4.79 & 3.60 & 1.31 & 1.14 \\
De Gerlache PSR* & 0.00 & 49.12 & 3.80 & 1.92 & 0.11 & 0.06 \\
\hline
\end{tabular}
\end{table}

\begin{table}
\centering
\caption{\textbf{Aspergillus survivability over extended time periods.} The table shows the percentage of survivable surface for \textit{Aspergillus} after X Earth Days during Spring/Autumn in different regions using a similar comparison as the preceding table. The * indicates that both direct and scattered fluxes have been modeled and accounted for.}
\vspace{4pt}
\label{tab:survivedays}
\begin{tabular}{lccccc}
\hline
\textbf{Site} & \textbf{0.5 Days} & \textbf{1 Day} & \textbf{3 Days} & \textbf{5 Days} & \textbf{7 Days} \\
\hline
Nobile Rim 2 & 22.57 & 12.51 & 6.06 & 4.44 & 3.68 \\
Connecting Ridge & 31.29 & 16.75 & 6.55 & 4.54 & 3.67 \\
De Gerlache Rim & 24.85 & 13.84 & 5.39 & 3.66 & 2.88 \\
De Gerlache PSR* & 47.68 & 40.19 & 22.81 & 12.98 & 4.57 \\
\hline
\end{tabular}
\end{table}

\end{document}